\documentclass[10pt,conference]{IEEEtran}
\IEEEoverridecommandlockouts
\usepackage{cite}
\usepackage{amsmath,amssymb,amsfonts}
\usepackage{algorithmic}
\usepackage{graphicx}
\usepackage{textcomp}
\usepackage{xcolor}
\usepackage{physics}
\usepackage{array}
\usepackage{comment}
\usepackage{subfloat}
\usepackage{subfig}
\usepackage{booktabs}
\usepackage{multirow}
\usepackage{cite}
\usepackage[utf8]{inputenc}
\usepackage[english]{babel}
\usepackage{url}
\usepackage{tabularx,booktabs}
\newcolumntype{C}{>{\centering\arraybackslash}X} 
\usepackage{lipsum} % filler text
\usepackage{amsmath}
\usepackage{graphicx}
\usepackage[colorlinks=true, allcolors=blue]{hyperref}
\usepackage{tabularx}
\usepackage{booktabs} % for \toprule, \midrule, \bottomrule
\usepackage[ruled,vlined]{algorithm2e}
\usepackage{cite}
\usepackage{url}
\usepackage{xurl}          % <-- fixes long URL overflow

\begin{document}
\title{Experimental Side Channel Analysis of Protocol Stages
in Quantum Identity Authentication

\author{
Marwan Elawady$^1$, Lance Young$^1$, Contessa Wilburn$^1$, Blaine Keyton$^1$, Carrie Houston$^1$,\\ Mohamed Shaban$^{1,2}$, and Muhammad Ismail$^1$
\vspace{1mm}
\\
{$^1$Cybersecurity Education, Research, and Outreach Center (CEROC) and Department of Computer Science,
}\\
{Tennessee Tech University, Cookeville, TN, USA.}\\
{$^2$Department of Mathematics, Faculty of Education, Alexandria University, Egypt} \\

Emails: \{mmelawady42, layoung43, crwilburn42, bckeyton42, cihouston42, mshaban, mismail\}@tntech.edu

\thanks{This work was supported by NSF Award \# 2322594.}
}
}

\maketitle
\begin{abstract}
Quantum networks can enable distributed computing and sensing. To realize these capabilities securely, quantum identity authentication is essential. Without authentication at the quantum layer, malicious repeaters may retain entanglement instead of performing swapping, enabling man‑in‑the‑middle attacks (MitM) between communicating parties. Authentication mitigates this threat by embedding authentication qubits within data qubits at positions and bases based on a secret key shared a priori. While prior work analyzes security and MitM detection guarantees, physical layer side channel analysis remains unexplored. If an attacker infers protocol stages, it can avoid authentication qubits and extract data qubits, rendering authentication ineffective. To this end, we carry out experimental studies using a quantum communication testbed. A beam splitter is used to tap a portion of the optical signal, allowing the observer to collect side channel data without disrupting the quantum state. We evaluate two sampling settings, where $30\%$ or $10\%$ of the signal is diverted. The collected side channel data includes photon arrival timing and optical power data obtained using a single-photon detector and a power meter. Using this dataset, we extract and engineer features that capture both timing dynamics and signal intensity variations. We then train machine learning models to classify protocol stages based solely on side channel observations. Our results show that protocol-stage inference is feasible with high accuracy, reaching $98\%$ (F1-score $97\%$) at $30\%$ sampling and $96\%$ (F1-score $94\%$) at $10\%$ sampling. These findings reveal an overlooked vulnerability and highlight the need for robust designs against side channel inference attacks.
\end{abstract}

\section{Introduction}
\IEEEPARstart{Q}{uantum} networks represent a major technological advance, enabling capabilities beyond classical communication systems. By supporting the transmission and processing of quantum information, quantum networks enable applications such as quantum key distribution (QKD), distributed quantum computing, and large-scale quantum sensing \cite{wehner2018quantum}. As these systems transition from laboratory prototypes to real-world deployments, understanding their security in practical settings becomes increasingly important. While quantum protocols offer strong theoretical security guarantees, these guarantees are typically established under idealized assumptions that overlook potential information leakage at the physical layer \cite{xu2020secure,dutta2022short}. In practical implementations, physical layer side channels may expose unintended information due to the interaction between protocol operations and the underlying hardware. In particular, different stages of a quantum protocol can inherently produce distinguishable physical signatures, such as variations in photon timing, optical power, and emission patterns, without requiring direct quantum state measurements. These signatures arise from how protocol stages utilize quantum resources and interact with the physical layer. Understanding such leakage is important, as it may reveal protocol behavior to unintended observers and impact both security and network operation. 

Quantum identity authentication (QIA) is a key protocol for securing quantum networks against malicious repeaters that retain entanglement rather than performing entanglement swapping, enabling man‑in‑the‑middle (MitM) attacks between communicating parties. QIA mitigates this threat by embedding authentication qubits within data qubits at secret positions and bases determined by a pre‑shared secret key, allowing receivers to verify communication integrity and detect active MitM attacks. Prior works have analyzed the security guarantees of QIA and its effectiveness in detecting MitM attacks. Yet, these analyses implicitly assume that authentication and data transmission stages are indistinguishable at the physical layer. If a passive observer can infer the QIA protocol stages, it can avoid interacting with the authentication qubits and selectively extract the data qubits. This enables information leakage without triggering detection, undermining the security guarantees of QIA by exposing stage‑dependent vulnerabilities. This raises a key question: can an external observer infer the execution stage of a QIA protocol based solely on physical layer side channel observations?

\subsection{Related Work}

Existing work on side channel analysis in quantum systems has primarily focused on QKD and quantum computing platforms, with an emphasis on extracting secret information or inferring low-level system behavior. In the QKD domain, several studies have demonstrated that physical layer emissions can leak sensitive information. For example, \cite{8539703} showed that a single electromagnetic trace can be sufficient to classify bit and basis choices, while \cite{Baliuka_2023} demonstrated that radio frequency emissions from QKD sender electronics can reveal substantial key information. Additional works have shown that imperfections in quantum sources and devices can lead to distinguishable physical emissions, enabling information extraction through indirect observations \cite{9531968}. 

Beyond QKD, side channel leakage has also been explored in quantum computing platforms. The work in \cite{Erata_2024} demonstrated that power side channels from quantum control hardware can be used to reconstruct executed quantum circuits. Similarly, \cite{10.1145/3716368.3735264} and \cite{11311002} investigated timing-based side channels in cloud-based and simulated quantum computing, showing that execution patterns and system characteristics can be inferred without direct access to quantum states. These efforts highlight that physical layer observations can reveal internal system behavior across different quantum platforms.

Despite these advances, existing approaches primarily focus on recovering secret information, such as key bits, or reconstructing computational processes under specific assumptions. These methods typically operate at the level of bit-level inference or circuit reconstruction, rather than identifying higher-level protocol execution stages. In particular, the problem of distinguishing between different operational phases within a QIA protocol based solely on passive observations of a communication channel remains largely unexplored. Understanding whether such observations can reveal stage-level behavior in QIA is critical for assessing potential information leakage and developing mechanisms to mitigate such leakage in emerging quantum networks.

\subsection{Contributions}
%Our results motivate new design requirements for quantum network protocols and architectures, specifically the need for resilience against structural side-channel inference attacks, achievable through mechanisms such as temporal obfuscation, optical power normalization, and protocol-stage difficult to differentiate techniques. To this end, our key contributions include:
To fill this gap, we develop and experimentally evaluate a framework for QIA protocol stage identification in quantum networks using physical layer side channel observations. The QIA protocol consists of distinct operational stages, namely data transmission and authentication. The goal is to infer these stages from physical layer side channel observations. To that end, we make the following contributions:
\begin{itemize}

\item We identify and experimentally validate a new class of side channel threats in quantum networks, where a passive adversary can infer the execution stage of a QIA protocol using only optical power and photon timing observations, without measuring transmitted data or disturbing the quantum channel.
\item We demonstrate the feasibility of QIA protocol stage inference on a real quantum communication testbed. The evaluation is conducted under different sampling/tapping ratios of $30\%$ and $10\%$, modeling varying levels of physical access (optical tapping).
\item We propose and evaluate a data-driven approach for QIA stage identification based on the collected side channel data. Our results achieve high classification accuracy of $98\%$ with F1-score $97$\% at $30\%$ sampling and $96\%$ with F1-score $94\%$ at $10\%$ sampling, demonstrating that QIA stages can be reliably inferred from passive observations.
\end{itemize}
The rest of this paper is organized as follows. Section \ref{sec:2} summarizes the QIA protocol and its operational principles. Section \ref{sec:3} describes the proposed methodology, including the problem formulation. Section \ref{sec:4} introduces the experimental setup and data collection process, followed by feature extraction, feature engineering, and the learning framework. Section \ref{sec:5} presents the experimental results and discusses the performance of the proposed approach. Finally, Section~\ref{sec:conclusion} concludes the paper and identifies some future directions.

\section{Background and Protocol Overview}
\label{sec:2}
QIA enables communicating parties to verify each other's identity during a quantum communication session. Several QIA protocols have been proposed in the literature, including schemes based on single photons, entangled states, counterfactual communication, and hybrid quantum-classical designs \cite{1,2,3,4,5,7,47,48,49}. These protocols typically perform authentication during dedicated phases or at the beginning of a communication session, ensuring that legitimate users are verified before data transmission proceeds. However, such approaches may still be vulnerable to evasion, as an adversary may avoid measuring authentication states and remain undetected during subsequent communication. A more robust approach is introduced in \cite{10757654}, where authentication is performed periodically by interleaving authentication rounds with data transmission. The timing and configuration of these authentication rounds are determined by a shared secret key, making their occurrence difficult to predict from a protocol-level perspective. Therefore, in this work, we implement the protocol proposed in \cite{10757654} to investigate its behavior at the physical layer.

During authentication, quantum states are prepared based on parameters derived from a shared secret key, including an encoding bit and a basis selection. The authentication qubit is prepared in one of four quantum states corresponding to the $Z$-basis $\{\ket{0}, \ket{1}\}$ and the $X$-basis $\{\ket{+}, \ket{-}\}$. The specific state is determined by the encoding value and the selected basis, which are derived from a segment of the prior shared key defined by the transfer length. The transfer length also determines the amount of data transmitted between consecutive authentication rounds. The authentication process is designed to preserve secrecy and avoid revealing information about the shared key through observable behavior. Thus, authentication rounds are embedded within data communication, making them indistinguishable at the protocol level under ideal conditions. The protocol, therefore, alternates between data transmission and authentication phases in a structured manner. The question we aim to answer is whether this structured behavior is reflected at the physical layer, where differences in state preparation, timing, and transmission patterns may arise across different stages. In this paper, we examine these stage-dependent characteristics to investigate whether protocol execution stages can be inferred through passive side channel observations.

\section{Problem Definition, Threat Model, and Experimental Testbed}
\label{sec:3}
In this section, we formally define the problem of inferring the QIA protocol stages. Then, we present the threat model, followed by a description of the experimental testbed used t facilitate our study. 
\subsection{Problem Definition}
We consider a quantum communication system in which two parties, Alice and Bob, execute a quantum protocol consisting of multiple operational stages, namely, data transmission and authentication. These stages occur sequentially during an active communication session and may exhibit different physical characteristics at the channel level. We consider the presence of a passive external observer with limited physical access (tapping) to the communication channel. The observer does not perform any direct quantum measurements on the transmitted states and does not interfere with the protocol execution. Instead, the observer can tap a fraction of the optical signal and collect physical layer information, including photon arrival timing and optical power measurements. Let $X$ denote the set of side channel observations collected over a given time interval, and let $S \in \{s_1, s_2\}$ denote the corresponding protocol stage during that interval, namely $s_1$ for data transmission and $s_2$ for authentication. The objective of the observer is to infer the protocol stage $S$ based solely on the observed side channel data $X$. Formally, the problem can be defined as learning a mapping
\begin{equation}
f: X \rightarrow S,
\end{equation}
where $f$ estimates the execution stage of the quantum protocol from passive physical layer observations. The goal is to determine whether such a mapping can be learned reliably under practical conditions, without access to protocol internals, such as shared key, transfer length, basis encoding, etc., or quantum state information.

\subsection{Threat Model}

We consider a passive external observer with limited physical access to the quantum communication channel between Alice (transmitter) and Bob (receiver). In the experimental testbed, the observer is positioned along the optical link connecting Alice and Bob, where a beam splitter is inserted to tap a fraction of the propagating optical signal.

Through its tapped port, the observer collects physical layer information, including photon arrival timing and optical power measurements, without accessing the transmitted quantum states directly. The observer does not perform any direct quantum measurement on the transmitted states and does not interfere with the protocol execution. As
such, the communication between legitimate parties remains functionally unaffected. The observer operates without introducing detectable disturbances to the quantum channel. The observer has no prior knowledge of the transmitted data, shared secret keys, or protocol parameters, and relies solely on side channel observations to infer protocol behavior.

\subsection{Experimental Testbed}
\label{sec:testbed}

\begin{figure*}
\centerline{\includegraphics[width=\linewidth]{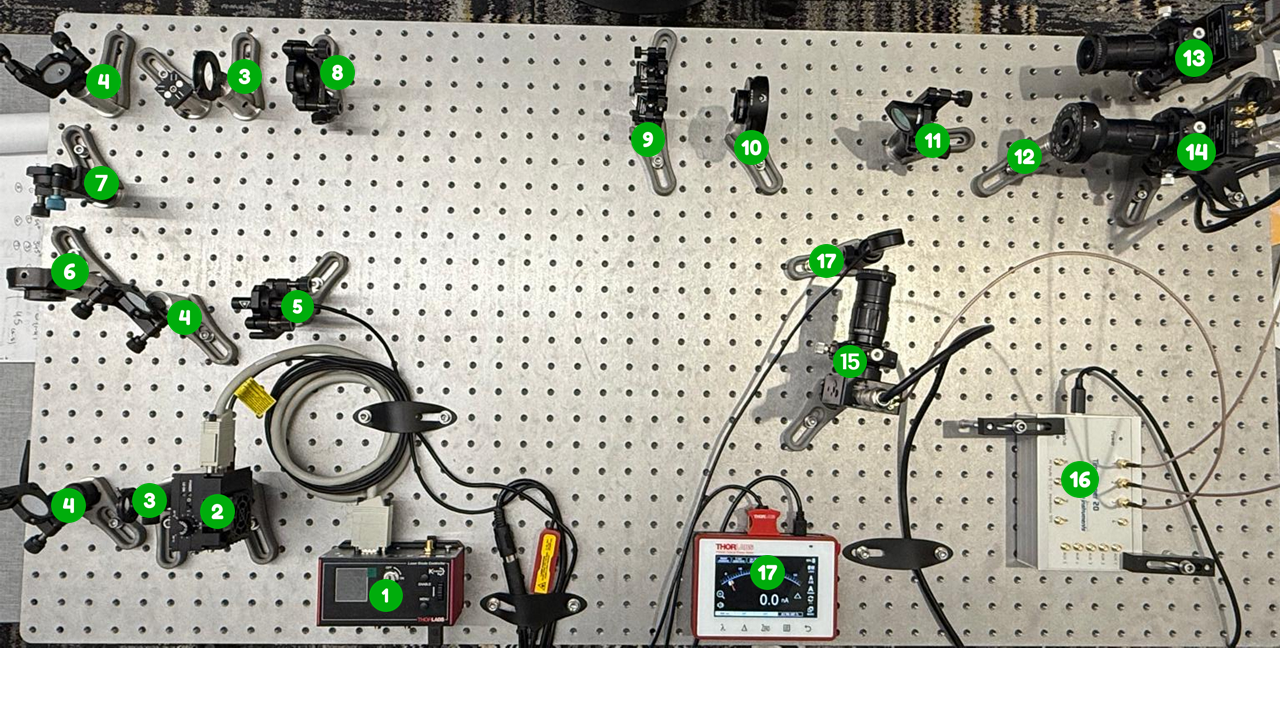}}
\vspace{-30pt}
\caption{(1) laser driver, (2) 405 nm pump laser source, (3) iris, (4) kinematic mirrors for beam alignment, (5) alignment laser, (6) 405 nm HWP, (7) temporal compensation crystal, (8) crossed BBO crystal for entangled photon generation, (9) spatial compensation crystals, (10) 808 nm HWP for state preparation and acts as Alice, (11) beam splitter for passive sampling, (12) linear polarizer, (13) heralding SPAD to enable coincedence-based filtering and temporal alignment, (14) SPAD Bob, (15) SPAD observer, (16) time tagger, (17) power meter that is used also as an observer. }
\label{fig:testbed}
\end{figure*}

To examine the proposed framework, we utilize a polarization entangled photon testbed that enables controlled execution of quantum communication protocols, as illustrated in Fig.~\ref{fig:testbed}. The setup is based on spontaneous parametric down conversion (SPDC), where a continuous wave pump laser operating at $405$ nm is used to generate pairs of polarization entangled photons through a beta barium borate (BBO) crystal. By properly adjusting the phase, the generated two-photon state can be expressed as
\begin{equation}
\ket{\psi} = \frac{1}{\sqrt{2}} \left(\ket{HH} + e^{i\phi} \ket{VV} \right),
\end{equation}
where $\phi$ is the relative phase controlled during the generation process, enabling the preparation of a maximally entangled Bell state. The generated photon pairs are separated into two optical paths, where one path is used for state preparation and measurement, while the other provides a heralding signal for timing and synchronization. The testbed supports the execution of the QIA workflow described in Section~\ref{sec:2}, including both data transmission and authentication phases. State preparation is achieved using polarization control components, including an $808$ nm half-wave plate (HWP), enabling the generation of quantum states in the $Z$ and $X$ bases. Photon detection is performed using single photon avalanche photodiodes (SPAD), while a time tagger records photon arrival times with high temporal resolution. To emulate the passive observer defined in the threat model, a beam splitter is introduced along the signal arm. The beam splitter is placed between the $808$ nm HWP, which acts as Alice, and the linear polarizer installed in front of the SPAD, which together represent Bob. Two sampling configurations are considered, where $30\%$ or $10\%$ of the optical signal is diverted to the observer, representing different levels of physical access. The tapped signal is directed to an observation unit that records physical layer information without affecting protocol execution. Specifically, the observer collects physical layer information using either a SPAD connected to the time tagger to record photon arrival timing or a power meter to measure optical power. Data is collected during repeated protocol executions, where authentication rounds are periodically interleaved with data transmission. This results in structured temporal behavior corresponding to different protocol stages. Importantly, the underlying hardware configuration remains unchanged across all experiments, ensuring that any observable differences arise from protocol execution rather than variations in the physical setup.

\section{Experimental Setup and Methodology}
\label{sec:4}
This section describes the QIA experimental implementation and the data collection and preprocessing process. It also presents the learning framework used for QIA protocol stage identification. 
\subsection{QIA Experimental Implementation}
\label{sec:qia_impl}
Polarization states are first calibrated to establish four reference states corresponding to $H$, $V$, $D$, and $A$. The horizontal and vertical states of the $Z$ basis are determined by identifying polarization settings that produce extremal coincidence counts. The diagonal states correspond to the $X$ basis and are defined relative to these references as
\begin{equation}
\ket{D} = \frac{1}{\sqrt{2}}(\ket{H} + \ket{V}), \quad
\ket{A} = \frac{1}{\sqrt{2}}(\ket{H} - \ket{V}).
\end{equation}
In implementation, this transformation is realized by setting the HWP to $22.5^\circ$, which maps the $Z$ basis states to the $X$ basis and effectively implements a Hadamard transformation up to a global phase. The QIA protocol is executed using a predefined shared secret key. The authentication state and the timing of authentication rounds are determined by parameters derived from the shared secret key, which control both the basis selection and state encoding. To introduce variability across experiments, multiple communication sessions are generated by permuting the key sequence, resulting in different protocol execution patterns across runs. Each session consists of a sequence of authentication rounds, where each round includes multiple data transmission slots followed by an authentication phase. During data transmission, the transmitted quantum states are selected randomly between two polarization states, $H$ and $V$, representing $\ket{0}$ and $\ket{1}$.

\subsection{Data Collection and Preprocessing}
\label{sec:data}

Data is collected using the experimental testbed described in Section~\ref{sec:testbed} under controlled and repeatable conditions. Multiple communication sessions are generated by permuting the shared secret key, resulting in different protocol execution patterns across runs. In addition, multiple experimental runs are conducted to capture variability in system behavior. The experiment is performed under different operating conditions by varying the pump laser current (e.g., $38$ mA and $43$ mA), with sufficient stabilization time between runs. Separate datasets are collected using different observation modalities, including photon arrival timing (via a time tagger) and optical power measurements (via a power meter). To evaluate the effect of side channel access, data is collected under different beam splitter configurations with tapping ratios of $30\%$ and $10\%$. The raw data is recorded as continuous measurements without explicit separation between protocol stages. 

To construct a structured dataset, a preprocessing step is applied to reorganize the data according to the protocol execution sequence. Specifically, data transmission segments are grouped and followed by their corresponding authentication segments, ensuring that the temporal ordering of protocol stages is preserved. The resulting dataset consists of labeled segments corresponding to data transmission and authentication phases, which are used for feature extraction and model training. 

Furthermore, normalization is applied to the features in the dataset to ensure that the learning process focuses on relative patterns in the data rather than absolute magnitudes that may vary across experiments. Outliers are mitigated using percentile-based clipping to improve robustness to measurement noise and transient fluctuations arising from detector effects (e.g., dark counts and afterpulsing) and optical background variations. The resulting feature values, $x$, are then normalized using min--max scaling, which maps each feature to the range $[0,1]$ according to
\begin{equation}
x_{\text{norm}} = \frac{x - x_{\min}}{x_{\max} - x_{\min}}.
\end{equation}

To address class imbalance between data transmission and authentication phases, we employ imbalance-aware techniques including Synthetic Minority Over-sampling Technique (SMOTE)~\cite{chawla2002smote}, balanced class weights, and stratified cross-validation. These methods ensure robust model training and reduce bias toward the majority class.

\subsection{Feature Extraction}
\label{sec:features}
To enable protocol stage identification, we construct feature representations from physical layer measurements collected by the external observer. These features are derived exclusively from the passive observation channel and are designed to capture both temporal dynamics and intensity variations associated with different execution stages.

\begin{itemize}
\item \textit{Time Tagger Features:}  
From photon arrival timestamps, we extract the photon count rate, defined as the number of detected photons per unit time. In addition, we compute inter-arrival time (IAT) statistics, including the mean, standard deviation, maximum value, and coefficient of variation. These features capture temporal variability and irregularity in photon arrivals, which reflect differences in state preparation and measurement behavior across protocol stages. To capture relative variations in signal strength, we also compute the normalized count rate by scaling the photon count rate with respect to its maximum value over the dataset. This feature highlights relative fluctuations in the photon detection rate, independent of absolute count levels.
\item \textit{Power Meter Features:}  
From the sampled optical signal, we extract optical power measurements in both milliwatts and decibel units (dBm), along with irradiance. These features characterize the magnitude and relative variation of the observed signal. Variations in these measurements reflect differences in optical intensity and stability between data transmission and authentication phases.
\end{itemize}
The selected features collectively capture both the magnitude and variability of the observed signals. Timing features describe photon arrival dynamics, while power based features characterize intensity fluctuations. Together, these features provide a comprehensive representation of stage-dependent physical layer behavior, enabling reliable differentiation between protocol execution phases.

\subsection{Feature Engineering}
In addition to the extracted features, we construct higher-level features to capture temporal structure and signal variability. Feature engineering is performed in four categories. First, interaction features are introduced to capture relationships between temporal and magnitude-based characteristics. Second, nonlinear transformations, including logarithmic and square-root operations, are applied to better represent skewed and heavy-tailed distributions. Third, rolling statistical features are computed over short time windows to capture local trends and stability. Fourth, lag and delta features are incorporated to model short-term temporal changes between consecutive samples. These engineered features enhance the representation of timing dynamics and signal variability, which are critical for distinguishing between protocol execution stages.

\subsubsection{Power Meter Feature Engineering}

For the power meter dataset, we construct engineered features to capture signal intensity, variability, and relationships between optical and electrical measurements. To model interactions between physical quantities, we include the power-current interaction, which reflects the joint behavior of optical power and the passive observer detector’s electrical response, highlighting deviations from their expected relationship. In addition, we introduce the irradiance variability-power interaction, which captures how noise level interacts with signal strength. To represent signal variability, we include both the irradiance coefficient of variation and the irradiance standard deviation, along with their normalized forms. These features provide both absolute and scale-invariant measures of fluctuations in the optical signal. Nonlinear transformations are applied to improve feature representation. Specifically, we use the square root of detector current to reduce the impact of large values while preserving relative differences. Overall, these engineered features focus on capturing intensity behavior, variability, and cross-feature relationships, enabling effective discrimination between protocol execution stages.

\subsubsection{Time Tagger Feature Engineering}

For the time tagger dataset, we construct engineered features to capture timing dynamics, variability, and relationships between photon arrival timing and detection rate. To model interactions between timing behavior and signal intensity, we introduce several cross-features. The count rate-IAT standard deviation interaction captures the relationship between detection rate and timing spread (the variability in photon arrival intervals). The count rate-IAT mean interaction captures the joint behavior of timing irregularity and spread. To represent the timing structure, we include the IAT regularity, defined as the inverse of the IAT coefficient of variation (CV) as follows
\begin{equation}
\text{IAT regularity} = \frac{1}{\mathrm{IAT}_{\text{cv}} + 0.01},
\end{equation} which emphasizes periods of uniform photon arrivals. We also include the count rate over IAT mean, which reflects the effective rate of photon arrivals relative to the average time gap between detections. Key timing features are retained, including the IAT coefficient of variation, the IAT mean, and the IAT standard deviation. These features directly characterize timing irregularity, average spacing, and spread of photon arrivals. Overall, these engineered features focus on capturing timing variability, timing structure, and interactions with detection rate, enabling effective discrimination between protocol execution stages.

\subsubsection{Combined Feature Engineering}

For the combined dataset, we construct engineered features that capture interactions between signal intensity (power meter) and timing dynamics (time tagger), enabling joint modeling of optical power and photon arrival behavior. Interaction features play a central role in this representation. The power-IAT standard deviation interaction captures the relationship between signal intensity and timing spread, linking optical power with temporal fluctuations. In addition, we introduce the dBm minus log count rate, which captures the difference between optical power in the logarithmic scale and the logarithm of photon count rate, providing a joint representation of intensity and detection behavior. The irradiance variability-IAT variability interaction captures the joint behavior of intensity fluctuations and timing irregularity. Furthermore, the combined set includes the following timing features: the count rate-IAT variability, the IAT regularity, the IAT variability, the IAT mean, and the IAT standard deviation, which together characterize timing spread, irregularity, and average spacing between photon detections. Overall, these features jointly capture timing dynamics, signal intensity, and their interactions, enabling a richer representation that improves discrimination between protocol execution stages.

%Similarly, the count rate-IAT variability interaction reflects cases where high detection rates coincide with irregular photon arrival patterns. To represent timing structure, we include the IAT regularity, defined as the inverse of IAT variability, which highlights periods of uniform photon arrivals. We also retain key timing features, including the IAT variability, IAT mean, and the IAT standard deviation, which together characterize timing spread, irregularity, and average spacing between photon detections. In addition, we introduce the dBm minus log count rate, which captures the difference between optical power in the logarithmic scale and the logarithm of photon count rate, providing a joint representation of intensity and detection behavior. The irradiance variability-IAT variability interaction captures the joint behavior of intensity fluctuations and timing irregularity. Overall, these features jointly capture temporal dynamics, signal intensity, and their interactions, enabling a richer representation that improves discrimination between protocol execution stages.

\vspace{-2mm}

\subsection{Learning Framework}
\label{sec:ml}
%To evaluate protocol stage identification, we construct three machine learning pipelines based on different observation modalities. The first pipeline uses only time-tagger features, the second uses only power meter features, and the third combines both feature sets. This design enables assessment of the discriminative power of each modality individually and jointly. We train a set of supervised learning models to classify protocol execution stages (data transmission vs. authentication). Specifically, we employ Random Forest \cite{breiman2001random} as a classical learning model, along with Long Short-Term Memory (LSTM) \cite{hochreiter1997long} and Gated Recurrent Unit (GRU) \cite{chung2014empirical} networks to capture temporal dependencies in the data. This combination allows us to evaluate both static feature-based learning and sequence-based modeling. For the combined pipeline, cross-modal interaction features are introduced to capture relationships between time-tagger and power meter measurements, enabling the model to leverage complementary information across modalities. To address class imbalance between data transmission and authentication phases, we employ imbalance-aware techniques including Synthetic Minority Over-sampling Technique (SMOTE)~\cite{chawla2002smote}, balanced class weights, and stratified cross-validation. These methods ensure robust model training and reduce bias toward the majority class.

To evaluate protocol stage identification, we examine three machine learning pipelines based on different observation modalities. The first pipeline uses only time tagger features (described in Section IV.D.2), the second uses only power meter features (described in Section IV.D.1), and the third combines both feature sets (described in Section IV.D.3). This design enables assessment of the discriminative power of each modality individually and jointly. We train a set of supervised learning models to classify protocol execution stages (data transmission vs. authentication). As outlined in Section IV.B, before model training, the dataset is labeled into two classes corresponding to the QIA protocol execution stages. Label $s_1 = 0$ represents the data transmission stage, which is the majority class, while label $s_2 = 1$ represents the authentication stage, which is the minority class, with SMOTE applied to balance the two classes as described in Section IV.B. We examine three machine learning models using the three feature sets described above (i.e., in total $9$ cases are examined, 3 different feature sets for each examined machine learning model). The considered machine learning models are Random Forest \cite{breiman2001random} as a tree-based machine learning model, along with Long Short-Term Memory (LSTM) \cite{hochreiter1997long} and Gated Recurrent Unit (GRU) \cite{chung2014empirical} models to capture temporal dependencies in the data. These three models are applied consistently across all dataset configurations. This design ensures that each model is evaluated under the same conditions across different collected datasets, allowing a fair comparison of their performance. It also enables us to assess the effectiveness of each feature set independently and in combination, without assigning specific models to specific feature types. To improve model performance, we tune key hyperparameters based on the data characteristics. The dataset is split into $80\%$ for training and $20\%$ for testing. From the training set, $10\%$ is further used as a validation set for hyperparameter tuning and model selection. This results in an effective split of $72\%$ training, $8\%$ validation, and $20\%$ testing. For the Random Forest model, validation is additionally performed using stratified $5$-fold cross validation to ensure robust performance evaluation. The Random Forest model used $500$ trees with a minimum leaf size of $2$ to better capture minority-class patterns. For neural networks (LSTM and GRU models), we use a learning rate of $10^{-4}$ with dropout regularization of $0.4$ and early stopping to prevent overfitting. In addition, focal loss with $\gamma = 2$ is applied to improve sensitivity to the minority class, and the classification threshold is optimized based on the F1-score of the authentication class.

\begin{figure}[t]
\centering
\includegraphics[width=\linewidth]{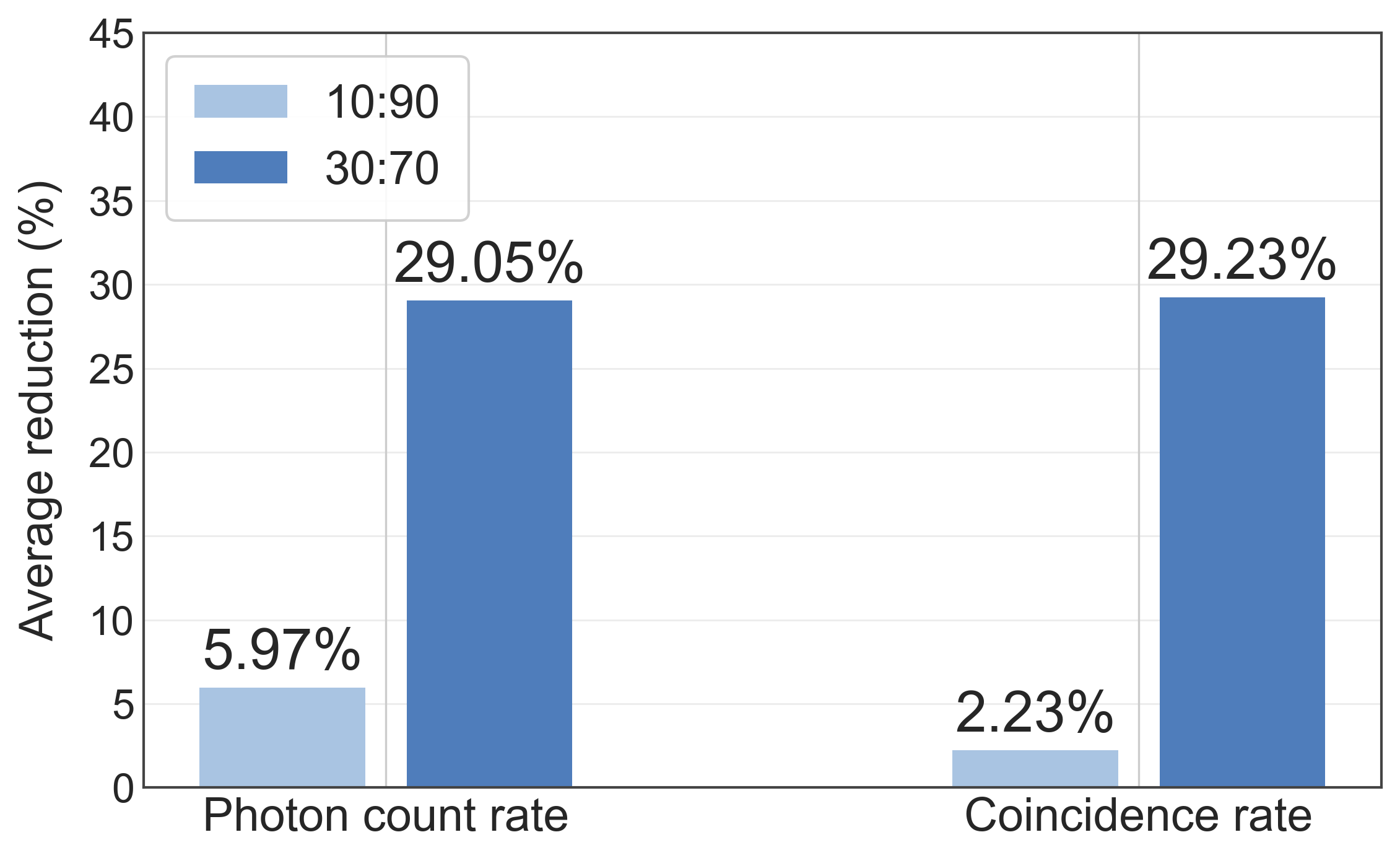}
\caption{Effect of passive sampling on signal photon count and coincidence rates. The $30{:}70$ configuration introduces significantly higher reductions in photon count rate and coincidence rate compared to the $10{:}90$ configuration.}
\label{fig:sampling_effect}
\end{figure}

\section{Experimental Results}
In this section, we evaluate the effectiveness of passive side channel observations for QIA protocol stage identification under different sampling ratios and feature configurations. The results include validation of quantum behavior under different passive samplings, classification performance across models and datasets, and analysis of feature importance. The classification results are obtained using datasets consisting of $9,000$ samples for each experimental configuration (i.e., for each machine learning model, tapping ratio, and feature set). 
\label{sec:5}
\subsection{Quantum Validation and Tapping Effects on Photon Count and Coincidence Rates}
We begin by validating that the passive observation mechanism does not alter the underlying quantum communication. To this end, we evaluate the CHSH Bell parameter (S-value) under different sampling configurations. In the absence of passive sampling, an S-value of $2.347 \pm 0.014$ is observed. With $10\%$ sampling, the S-value is $2.359 \pm 0.014$, while at $30\%$ sampling it is $2.324 \pm 0.016$. All values remain above the classical bound of $2$, confirming the presence of entanglement across all configurations. These results indicate that extracting side channel information through passive sampling does not disrupt the observed non-classical correlations. This validation supports the hypothesis that protocol behavior can be observed at the physical layer without collapsing the quantum state.

\begin{table*}[t]
\renewcommand{\arraystretch}{1.3}
\centering
\caption{Classification Performance at $38$ mA}
\label{tab:38ma}
\begin{tabular}{|l|cc|cc|cc|cc|cc|cc|}
\hline
\multirow{3}{*}{\textbf{Model}} 
& \multicolumn{4}{c|}{\textbf{Time Tagger}} 
& \multicolumn{4}{c|}{\textbf{Power Meter}} 
& \multicolumn{4}{c|}{\textbf{Combined}} \\ \cline{2-13}

& \multicolumn{2}{c|}{\textbf{10\%}} & \multicolumn{2}{c|}{\textbf{30\%}}
& \multicolumn{2}{c|}{\textbf{10\%}} & \multicolumn{2}{c|}{\textbf{30\%}}
& \multicolumn{2}{c|}{\textbf{10\%}} & \multicolumn{2}{c|}{\textbf{30\%}} \\ \cline{2-13}

& Acc. & F1-Score & Acc. & F1-Score & Acc. & F1-Score & Acc. & F1-Score & Acc. & F1-Score & Acc. & F1-Score \\ \hline

LSTM          
& $76.27$ & $65.89$ & $87.05$ & $78.15$ 
& $68.66$ & $63.08$ & $75.27$ & $68.86$ 
& $82.00$ & $74.14$ & $84.38$ & $78.62$ \\ \hline

GRU           
& $78.05$ & $66.89$ & $87.33$ & $77.61$
& $80.05$ & $69.79$ & $80.22$ & $71.69$ 
& $80.11$ & $73.36 $& $86.50 $& $81.14$ \\ \hline

Random Forest 
& $90.18$ & $83.18$ & $92.77$ & $87.94$ 
& $89.41$ & $83.05$ & $90.51$ & $85.80$ 
& $93.00$ & $89.16$ & $95.23$ & $92.60$ \\ \hline

\end{tabular}
\end{table*}

\begin{table*}[t]
\renewcommand{\arraystretch}{1.3}
\centering
\caption{Classification Performance at $43$ mA}
\label{tab:43ma}
\begin{tabular}{|l|cc|cc|cc|cc|cc|cc|}
\hline
\multirow{3}{*}{\textbf{Model}} 
& \multicolumn{4}{c|}{\textbf{Time Tagger}} 
& \multicolumn{4}{c|}{\textbf{Power Meter}} 
& \multicolumn{4}{c|}{\textbf{Combined}} \\ \cline{2-13}

& \multicolumn{2}{c|}{\textbf{10\%}} & \multicolumn{2}{c|}{\textbf{30\%}}
& \multicolumn{2}{c|}{\textbf{10\%}} & \multicolumn{2}{c|}{\textbf{30\%}}
& \multicolumn{2}{c|}{\textbf{10\%}} & \multicolumn{2}{c|}{\textbf{30\%}} \\ \cline{2-13}

& Acc. & F1-Score & Acc. & F1-Score & Acc. & F1-Score & Acc. & F1-Score & Acc. & F1-Score & Acc. & F1-Score \\ \hline

LSTM          
&$ 87.41$ & $76.24$ & $90.77$ & $83.44$ 
& $87.88$ & $79.68$ & $72.00$ & $67.37$ 
& $90.27$ & $85.30$ & $94.11$ & $90.76$ \\ \hline

GRU           
& $87.00$ & $77.65$ & $89.72$ & $82.81$ 
& $88.38$ & $80.98$ & $77.11$ & $68.57$ 
& $91.88$ & $86.61$ & $94.47$ & $91.80$ \\ \hline

Random Forest 
& $91.95$ & $86.38$ & $93.56$ & $88.67$ 
& $92.68$ & $88.47$ & $94.60$ & $91.56$ 
& $96.32$ & $94.21$ & $98.10$ & $97.03$ \\ \hline

\end{tabular}
\end{table*}

To examine the effect of sampling on the observed signals, we quantify the impact of passive sampling on photon count and coincidence rates at the physical layer. As illustrated in Fig.~\ref{fig:sampling_effect}, a sampling ratio of $30$\% leads to a reduction of approximately $29.05$\% in photon count rate and $29.23$\% in coincidence rate. In contrast, a $10$\% sampling ratio results in smaller reductions of approximately $5.97$\% and $2.23$\%, respectively. These results show that higher sampling ratios lead to larger reductions in the measured photon count and coincidence rates. However, the underlying non-classical correlations such as entanglement are preserved, indicating that passive sampling does not alter the quantum behavior of the system. This behavior reflects an inherent tradeoff between observability and rate such that higher sampling ratios expose more side channel information but introduce greater reduction in photon count and coincidence rates, whereas lower sampling ratios limit such reduction while still retaining sufficient information for analysis, as we will show next.

%\vspace{-4mm}

\subsection{QIA Protocol Stage Identification Results}
Table~\ref{tab:38ma} summarizes the classification accuracy across different models, sampling ratios, and observation modalities, demonstrating that protocol execution stages can be reliably distinguished using passive physical layer observations. A clear distinction is observed between the two types of features. Features derived from photon arrival measurements consistently achieve higher accuracy than those based on optical power measurements across most configurations. This behavior is attributed to the nature of protocol execution, where transitions between different operational stages introduce variations in photon detection patterns. These variations are reflected in photon arrival statistics, enabling timing features to capture distinctive signatures associated with each stage. In contrast, optical power measurements represent the aggregate signal magnitude, which tends to vary less consistently across protocol stages, limiting their ability to differentiate protocol behavior with high accuracy compared to the timing features. Nevertheless, power-based features capture coarse variations in signal magnitude and provide complementary information. This complementary behavior becomes evident when both feature types are combined. The integration of photon arrival and power-based features leads to improved classification performance, exceeding $90\%$ at $10\%$ sampling and reaching up to $95\%$ at $30\%$. This indicates that timing features capture the dominant stage-dependent characteristics, while power measurements provide additional contextual information that enhances overall discrimination. The combined representation enables the model to capture both fine-grained timing dynamics and broader signal variations.

\begin{figure*}
\centering
\subfloat[]{\includegraphics[width= 2.2in]{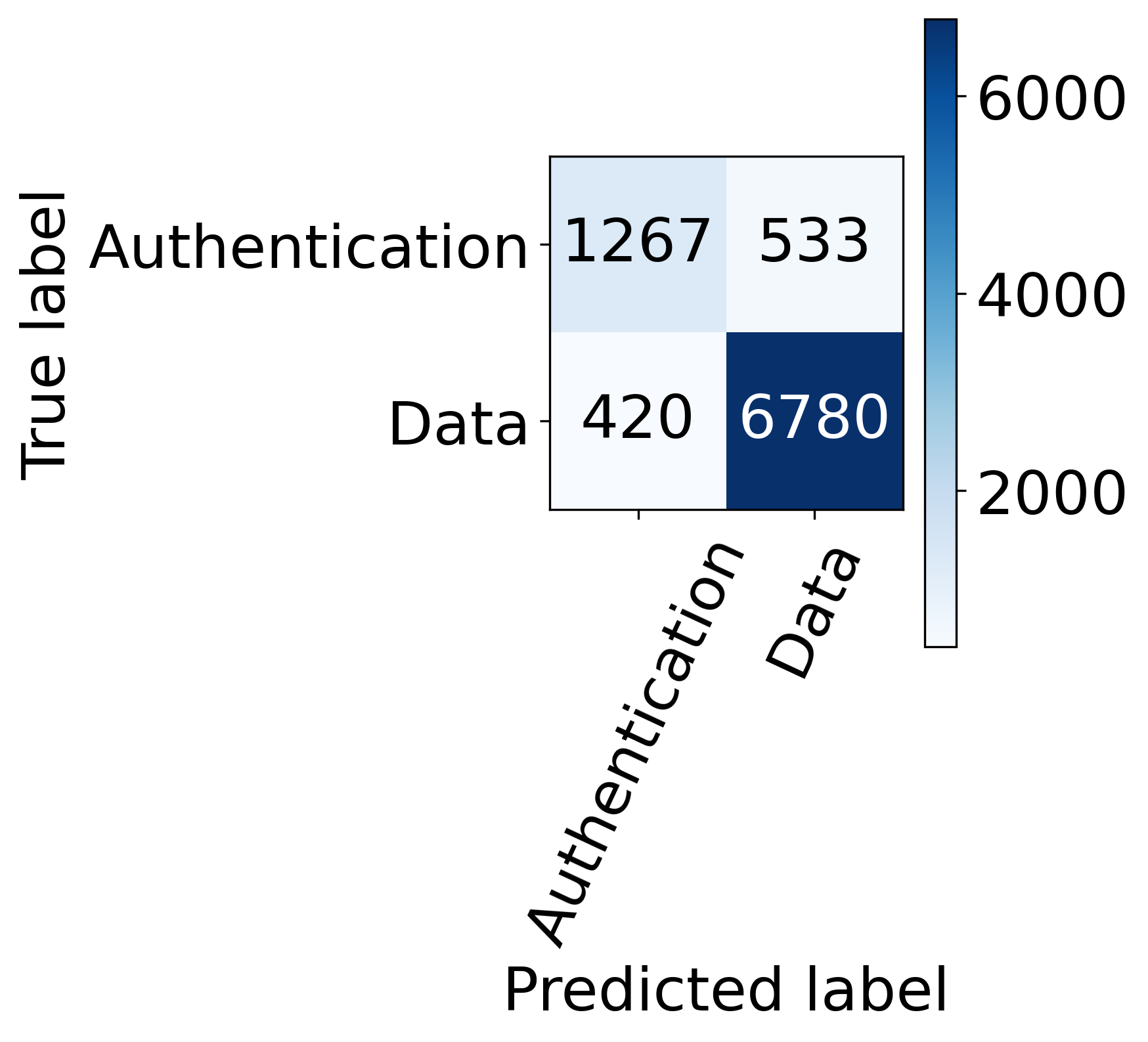}\label{fig:confusion_10_90_PM_38}} \;\;
\subfloat[]{\includegraphics[width= 2.2in]{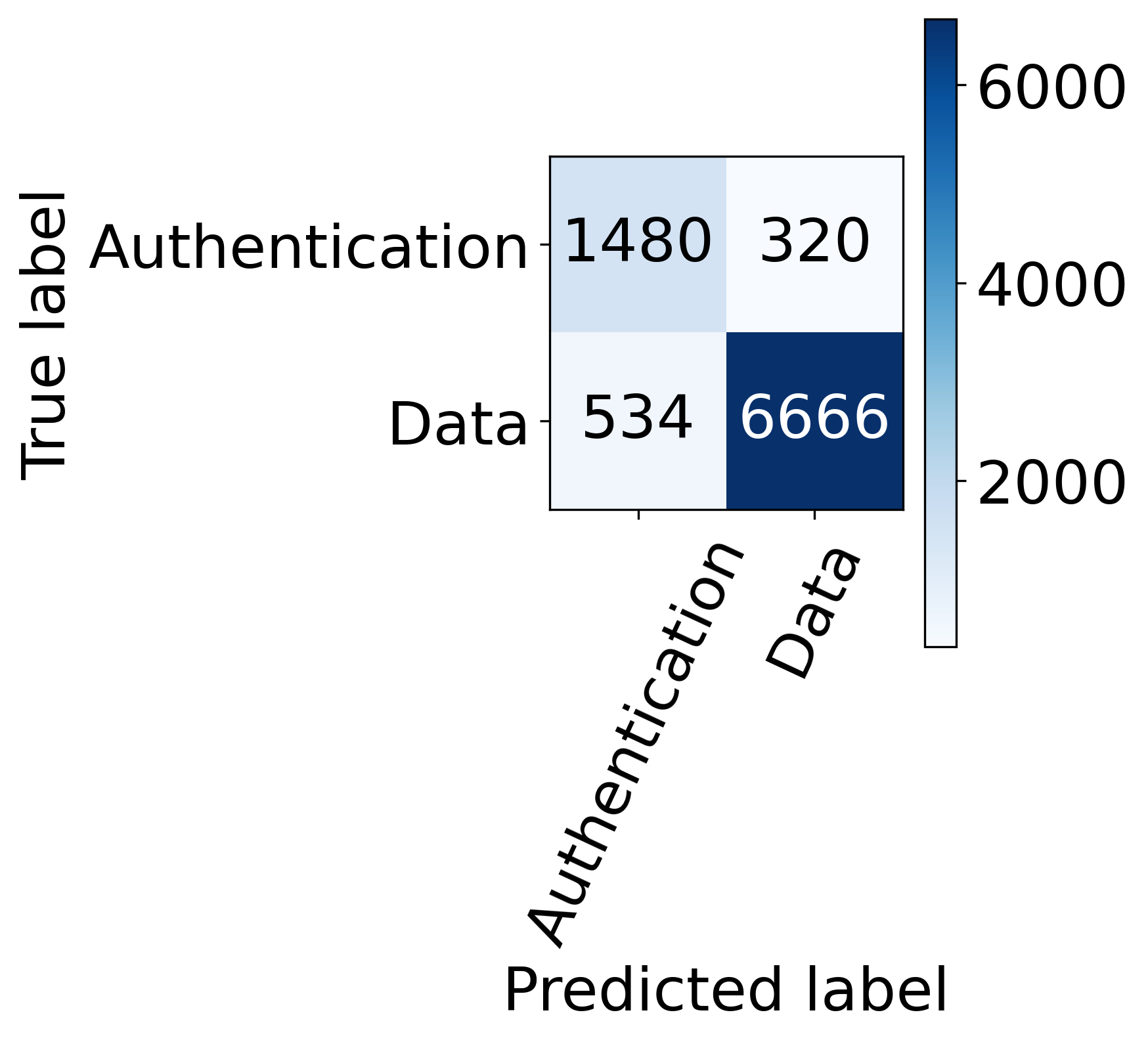}\label{fig:confusion_30_70_PM_38}} \;\;
\subfloat[]{\includegraphics[width= 2.2in]{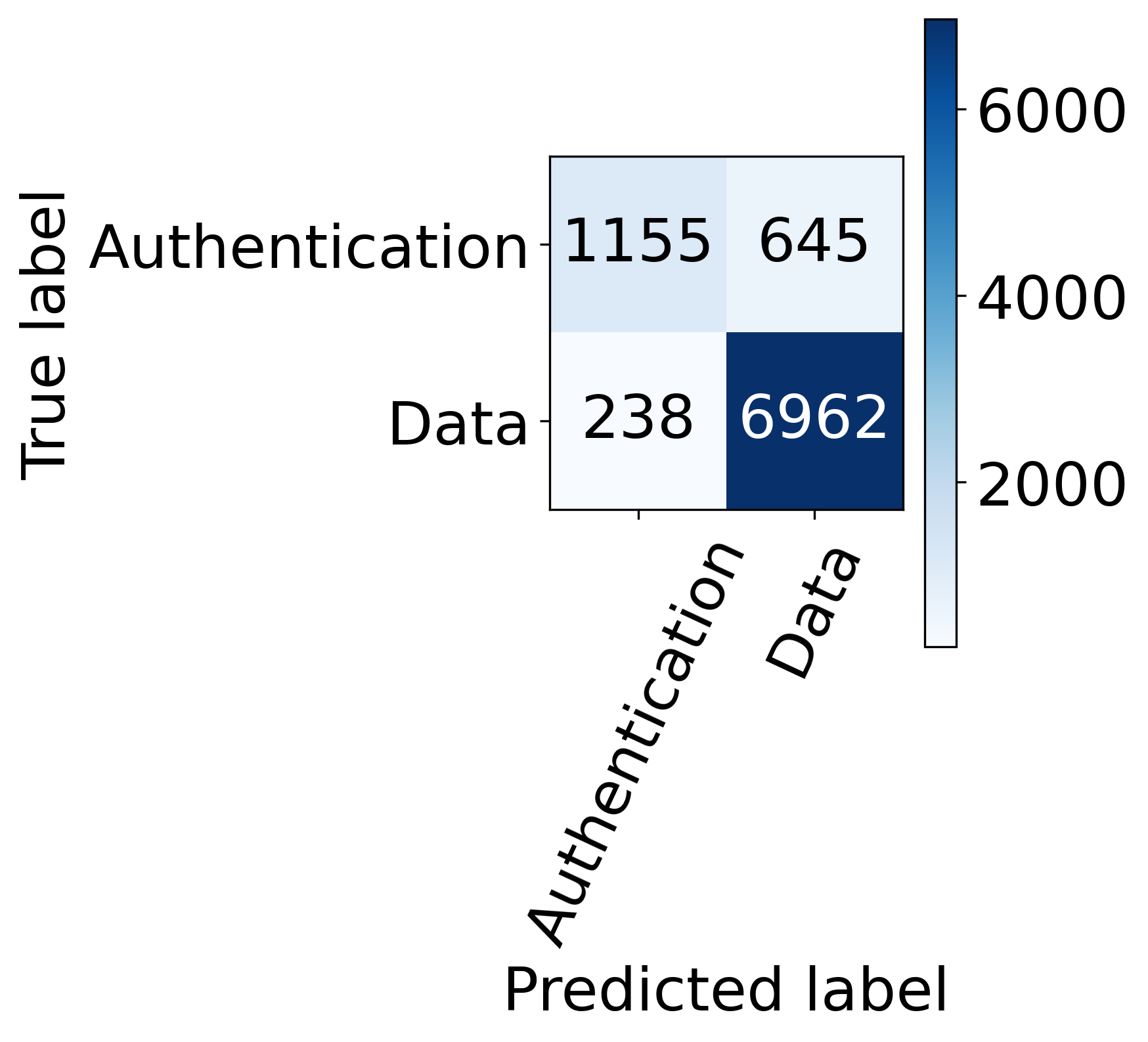}\label{fig:confusion_10_90_tt_38}} \;\;
\subfloat[]{\includegraphics[width= 2.2in]{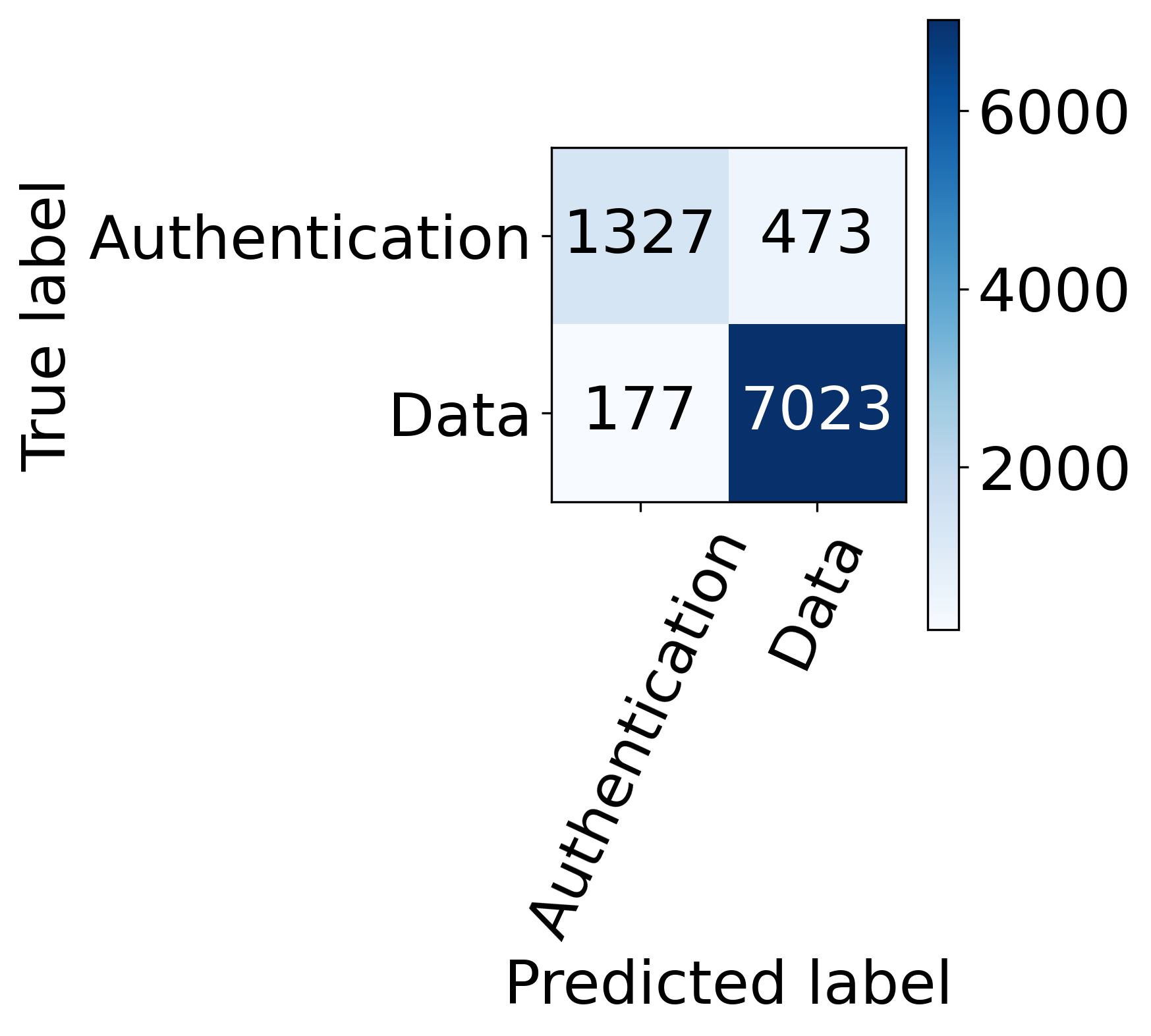}\label{fig:confusion_30_70_tt_38}} \;\;
\subfloat[]{\includegraphics[width= 2.2in]{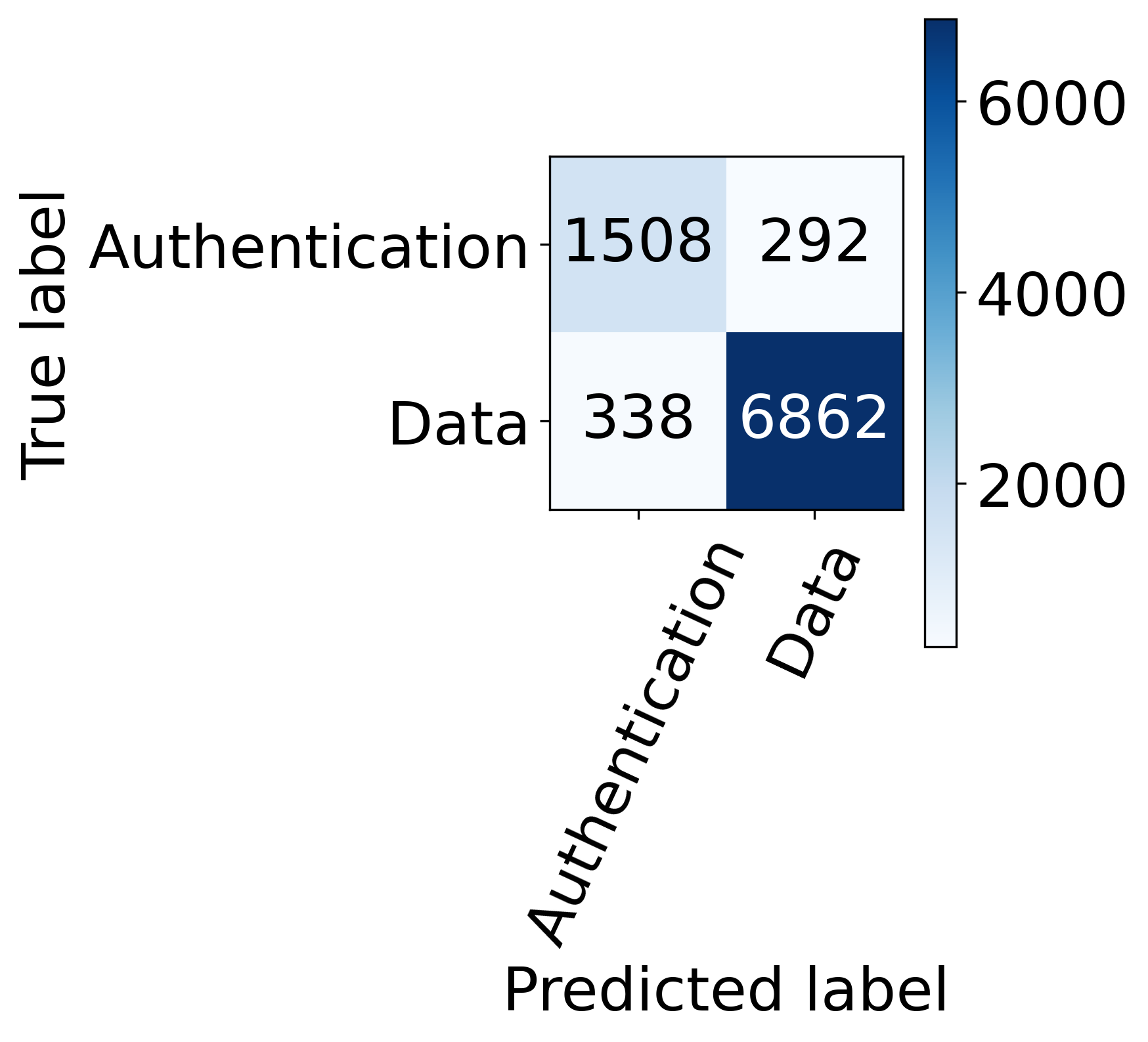}\label{fig:confusion_10_90_combined38}}\;\;
\subfloat[]{\includegraphics[width= 2.2in]{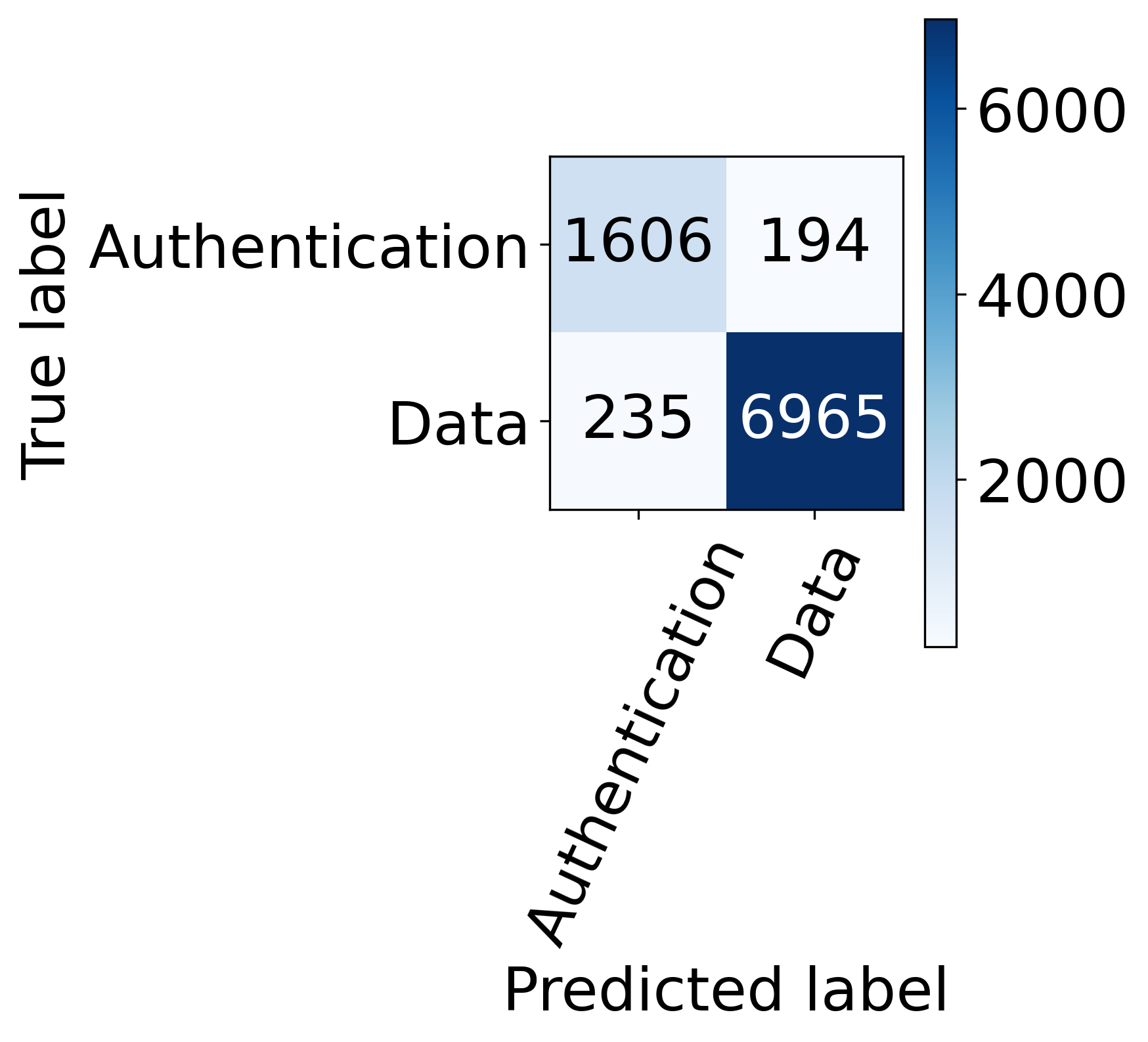}\label{fig:confusion_30_70_combined38}}
\caption{Confusion matrices illustrating the impact of feature modality and sampling ratio on protocol classification. (a) power meter at $10$\%, (b) power meter at $30$\%, (c) time tagger at $10$\%, (d) time tagger at $30$\%, (e) combined at $10$\%, and (f) combined at $30$\%. A higher sampling ratio of $30$\% leads to better class separability compared to $10$\%, while feature fusion (power meter + time tagger) consistently improves classification performance.}
\label{fig:matrices}
\end{figure*}

The impact of model selection is also consistent across configurations. Random Forest achieves higher accuracy compared to sequence-based models, indicating that the engineered features already capture the relevant temporal structure. This suggests that explicit feature design effectively encodes stage-dependent behavior, reducing the need for more complex sequential models such as LSTM and GRU.

To gain insight into the impact of side channel access on stage inference, we evaluate the effect of sampling ratio on classification performance. Increasing the sampling ratio from $10\%$ to $30\%$ improves classification performance due to increased visibility of the underlying signal. However, the ability to maintain high accuracy even at $10\%$ sampling demonstrates that protocol-stage information remains observable under constrained access conditions. It is important to note that higher sampling ratios introduce greater rate reductions to the measured signal, as discussed in the reduction in photon count and coincidence rates analysis. While higher sampling ratios provide improved classification performance, such configurations may not be suitable for low-impact or stealthy observation scenarios, as shown in Fig.~\ref{fig:sampling_effect}. In contrast, lower sampling ratios maintain minimal reduction in photon count and coincidence rates (as in Fig.~\ref{fig:sampling_effect}) while still enabling reliable stage inference (accuracy $90 - 93\%$ and F1-score $83 - 89\%$), highlighting a practical tradeoff between observability and rate reductions.

Table~\ref{tab:43ma} confirms that the same trends persist under higher laser power. While overall identification performance improves, reaching up to $98\%$ with F1-score $97$\% for the combined feature set at $30\%$ tapping ratio and $96\%$ with F1-score $94$\% for the combined feature set at $10\%$ tapping ratio, the relative behavior across features, models, and sampling ratios remains consistent with the performance shown in Table~\ref{tab:38ma}. This consistency indicates that the observed side channel leakage is driven by inherent protocol behavior rather than specific operating conditions.

The confusion matrices in Fig.~\ref{fig:matrices} provide further insight into classification behavior across feature sets and sampling configurations. Under the $30$\% sampling ratio, both time tagger and power meter features exhibit strong separation between protocol stages, with limited misclassification. This indicates that increased observability improves the reliability of stage dependent patterns in both time tagger and power-based measurements. In addition, the minority class is more accurately classified at $30$\% sampling compared to $10$\%, as reflected by the reduced number of misclassified authentication samples. The time tagger features also show slightly improved classification of the majority class, indicating a marginal advantage in capturing dominant time tagger patterns. When both feature types are combined, the level of misclassification is further reduced, indicating that the complementary information provided by time tagger and power-based features helps resolve ambiguities that remain when each modality is used independently. At the $10$\% sampling ratio, both feature modalities exhibit increased confusion between stages due to reduced observability of the underlying signal. Time tagger features show slightly higher sensitivity to this reduction, reflecting the stochastic nature of photon arrivals under constrained sampling. Power-based features maintain comparable separability, although minor misclassification persists due to the limited variation in aggregate signal magnitude at lower sampling levels. The combined feature set mitigates these effects by leveraging complementary information from both time tagger and power meter features. This results in improved discrimination between protocol stages, particularly in scenarios where individual feature sets are insufficient.

\subsection{Feature Importance Analysis}

%\vspace{-2mm}

To interpret the contribution of engineered features, we analyze permutation feature importance \cite{altmann2010permutation} for the best performing Random Forest model across different datasets and sampling configurations. In total, $16$ unique features are constructed, including both raw and engineered features.

\subsubsection{Power Meter Feature Analysis}

\begin{figure*}
\centering
\subfloat[]{\includegraphics[width=3.5in]{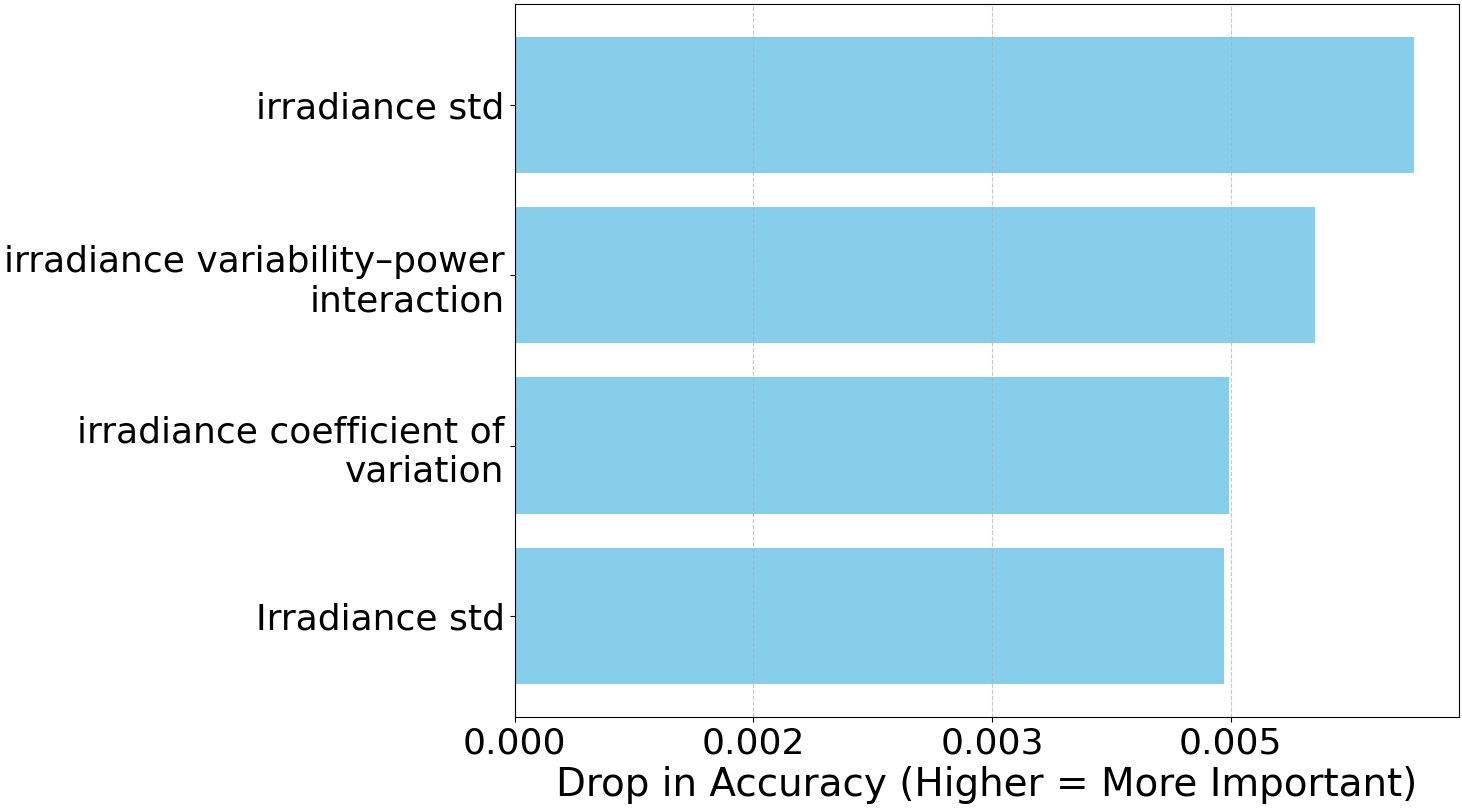}\label{fig:PFI_PM1090}}
\subfloat[]{\includegraphics[width=3.5in]{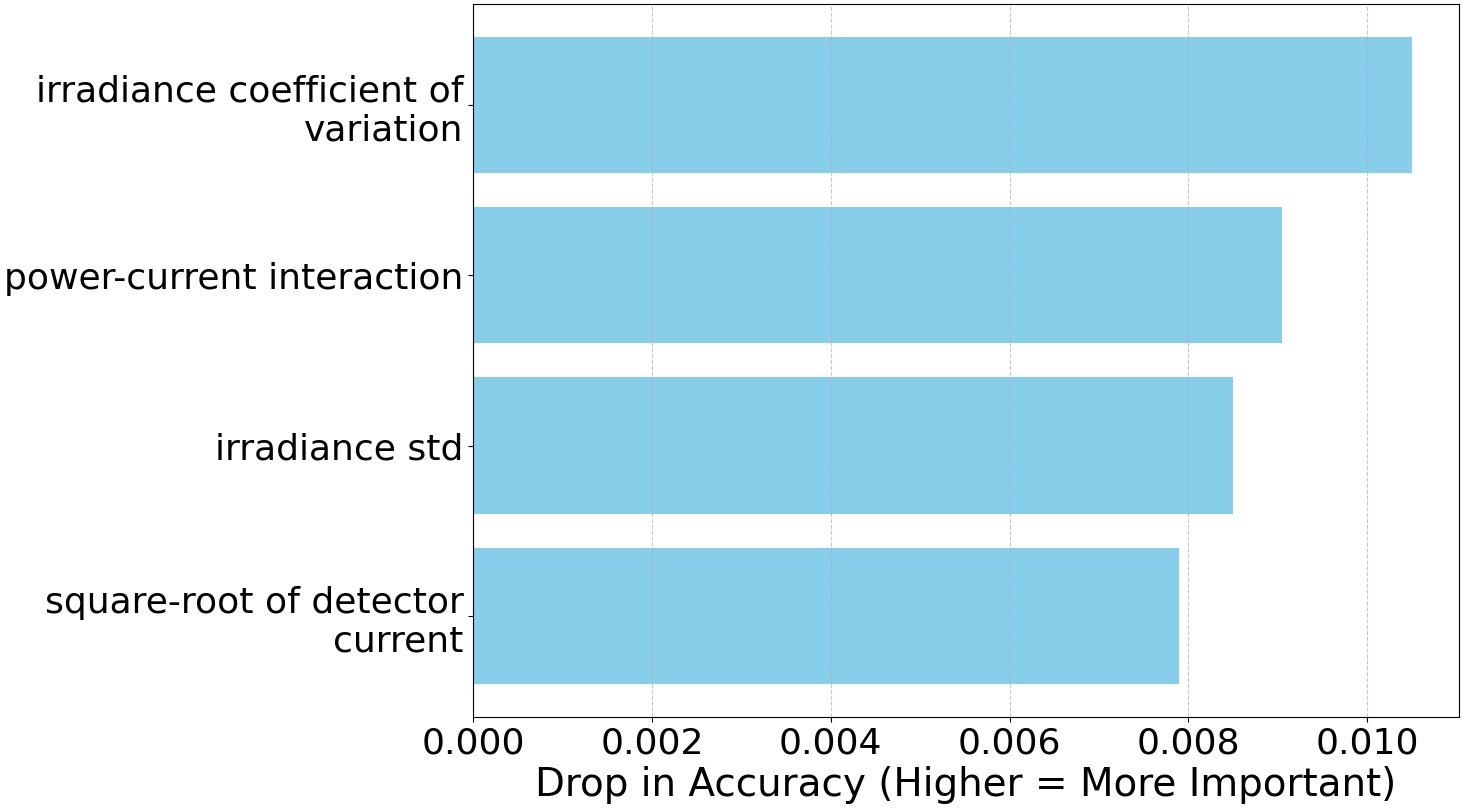}\label{fig:PFI_PM3070}}
\caption{Feature importance for the power meter dataset: (a) $10$\%, (b) $30$\%.}
%\vspace{-5mm}
\label{fig:pfi_pm}
\end{figure*}
\begin{figure*}
\centering
\subfloat[]{\includegraphics[width=3.5in]{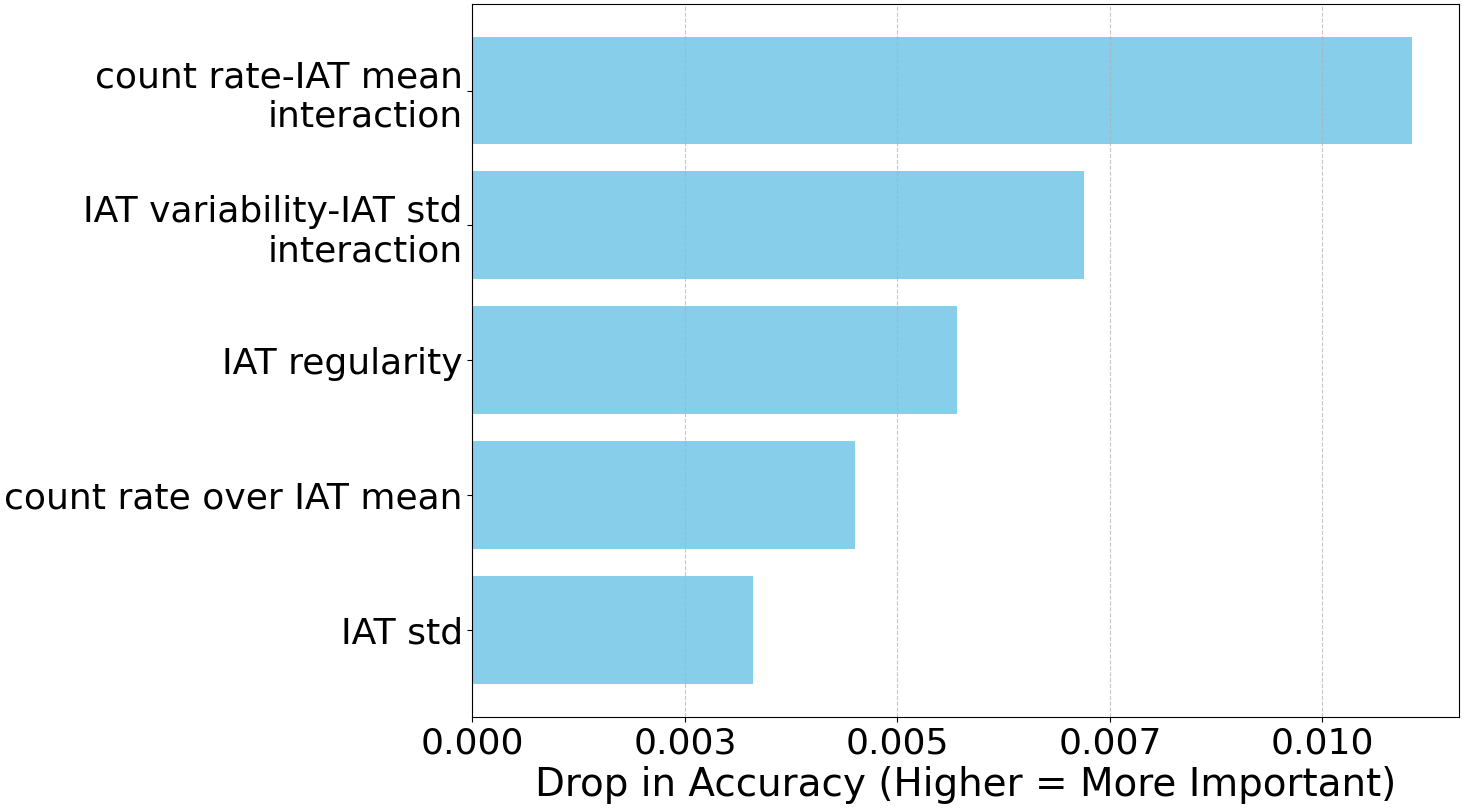}\label{fig:PFI_TT1090}}
\subfloat[]{\includegraphics[width=3.5in]{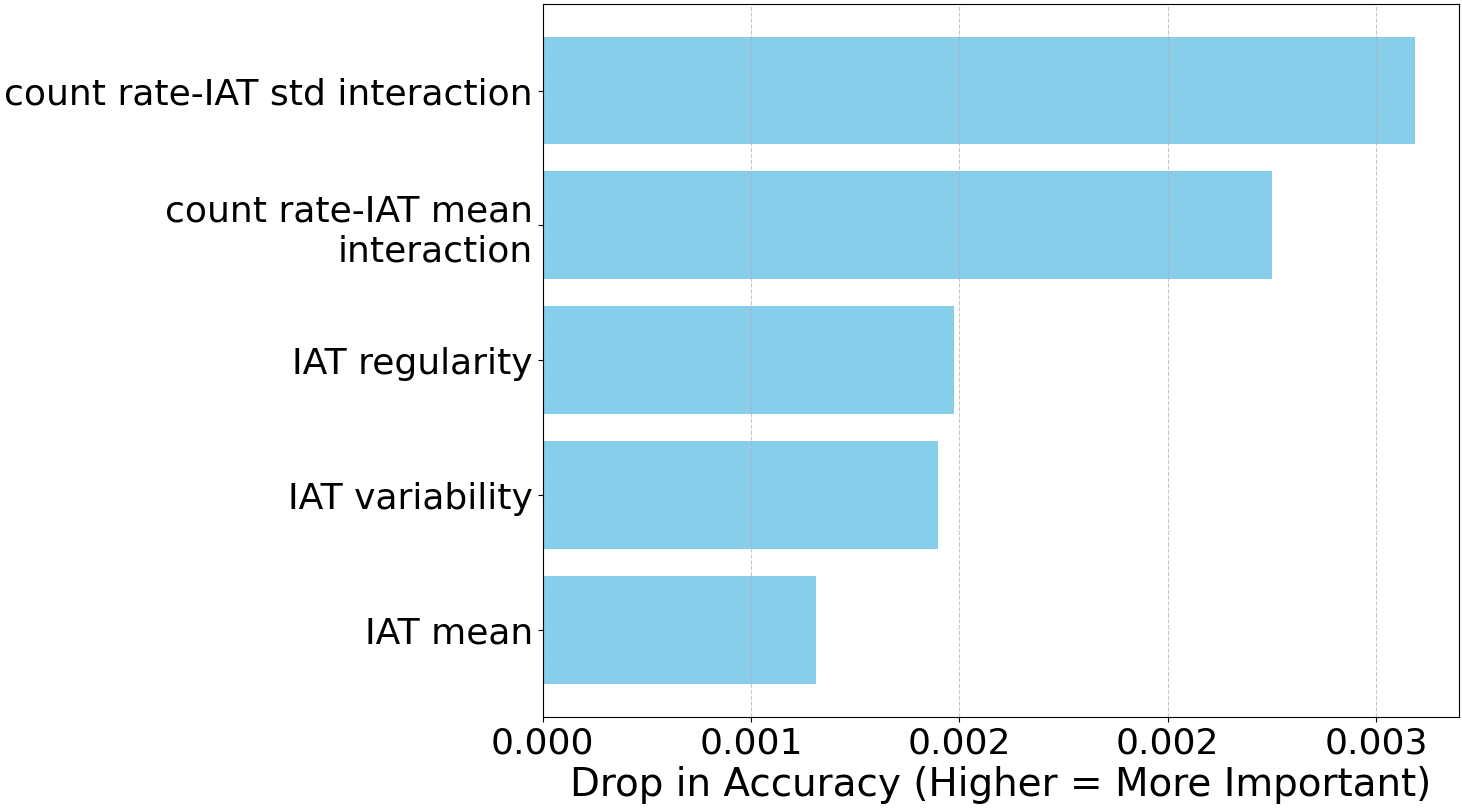}\label{fig:PFI_TT3070}}
\caption{Feature importance for the time tagger dataset: (a) $10$\%, (b) $30$\%.}
%\vspace{-5mm}
\label{fig:pfi_tt}
\end{figure*}
\begin{figure*}
\centering
\subfloat[]{\includegraphics[width=3.5in]{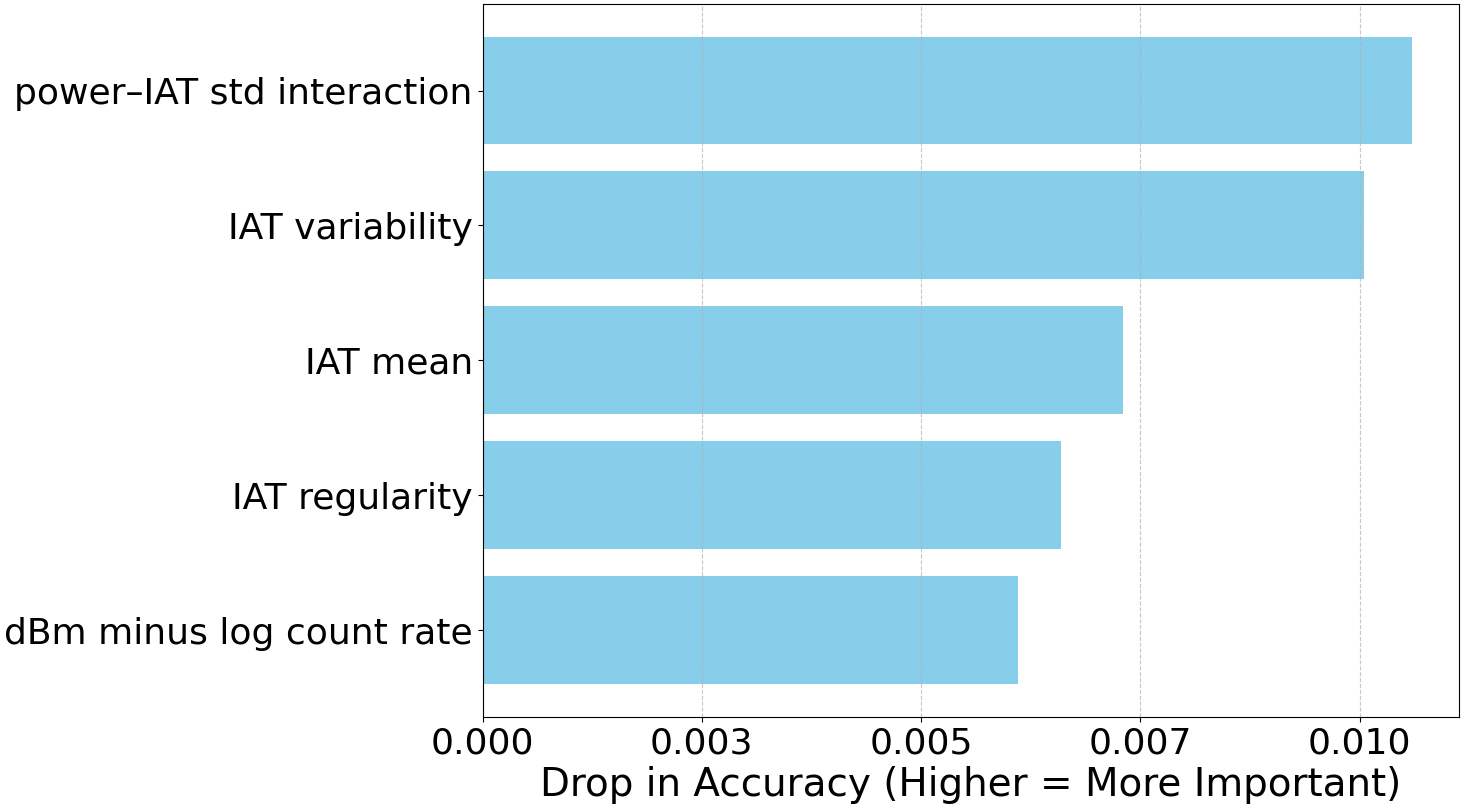}\label{fig:PFI_combined1090}}
\subfloat[]{\includegraphics[width=3.5in]{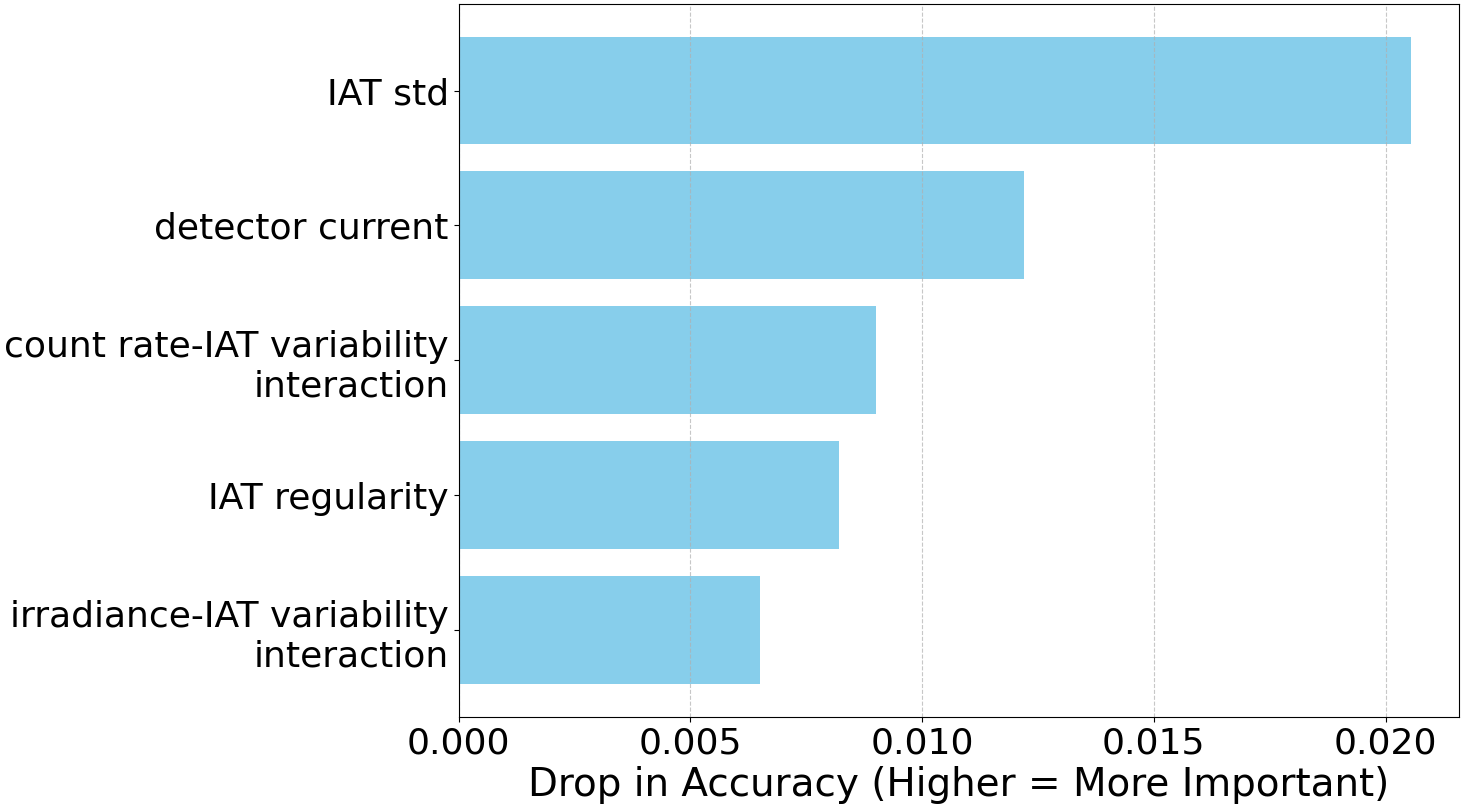}\label{fig:PFI_combined3070}}
\caption{Feature importance for the combined dataset: (a) $10$\%, (b) $30$\%.}
%\vspace{-5mm}
\label{fig:pfi_combined}
\end{figure*}

Figs.~\ref{fig:PFI_PM1090} and \ref{fig:PFI_PM3070} show the feature importance results for the power meter dataset under the $10$\% and $30$\% configurations, respectively. In both cases, features capturing variations in signal intensity and interactions between measurements provide the highest contribution. At the $10$\% configuration, feature importance is more distributed, with multiple features contributing moderately to classification. In contrast, under the $30$\% configuration, clearer patterns emerge, and a smaller subset of features becomes dominant. In particular, irradiance variability and interaction-based features show increased importance. Overall, these results indicate that power meter features capture protocol-specific information primarily through signal variability and cross-feature relationships. In contrast, raw intensity measurements without power or interaction-based features provide limited discrimination.

\subsubsection{Time Tagger Feature Analysis}

Figs.~\ref{fig:PFI_TT1090} and \ref{fig:PFI_TT3070} present the feature importance results for the time tagger dataset. Features such as IAT variability, IAT standard deviation, and their interaction with count rate provide the strongest contribution. These features capture irregularities and fluctuations in photon arrival patterns, which directly reflect protocol stage behavior. Compared to the power meter case, feature importance is more concentrated on a small set of dominant features, indicating that timing dynamics provide a stronger and more direct source of side channel information. Overall, these results confirm that timing-based features are highly effective for protocol stage identification due to their ability to capture dynamic behavior in photon arrivals at different protocol stages.

\subsubsection{Combined Power Meter and Time Tagger Feature Analysis}

Figs.~\ref{fig:PFI_combined1090} and \ref{fig:PFI_combined3070} show the feature importance results when combining power meter and time tagger features. In both $30$\% and $10$\% configurations, interaction-based features dominate the model. In particular, features that couple signal intensity with timing variability, such as power-IAT interactions and count rate-IAT variability interactions, provide the highest contribution. Time tagger features remain strong contributors, while intensity-based features provide complementary information. This combination leads to a more balanced and informative representation of the system. Compared to individual feature sets (power only and time tagger only), the combined feature set results in clearer feature dominance and improved separability, especially under the $30$\% configuration. This explains the higher classification performance observed for the combined model. Overall, these results demonstrate that protocol stage leakage is best captured when both timing dynamics and signal intensity are jointly modeled, with interaction features playing a central role.

%\vspace{-2mm}

\subsection{Discussion}

The experimental results demonstrate that a passive observer can identify the randomly interleaved authentication and data transmission phases with high accuracy solely by observing the quantum communication link between Alice and Bob and without any prior access to the shared secret key. These findings are possible with low impact on the communication link in terms of photon count and coincide rates at low sampling/tapping level of $10\%$ and without disturbing/collapsing the quantum states. These findings mean that malicious repeaters in the middle between Alice and Bob can passively observe power and timing features to infer whether the current transmission belongs to the data or the authentication stage. As such, data transmissions can be extracted while keeping the authentication transmission untouched, rendering the QIA ineffective and making undetectable MitM attacks feasible.

\section{Conclusion and Future Work}
\label{sec:conclusion}
This paper investigated the feasibility of inferring QIA protocol execution stages in quantum communication systems using passive physical layer observations. Through experimental evaluation on a polarization-entangled photon testbed, we demonstrated that stage-dependent behavior can be reliably identified using photon arrival statistics and optical power measurements, without performing any direct quantum state measurement. The results show that timing features derived from photon detection patterns provide strong discriminative capability, while power-based features offer complementary information that enhances classification performance when combined. Importantly, high accuracy is achieved even under constrained sampling conditions, indicating that protocol stage information remains observable with limited access to the optical signal. This highlights the need to consider physical layer leakage when designing secure and robust quantum communication systems. Future work will investigate the impact of noise and channel impairments on protocol stage inference, as well as develop techniques to mitigate such leakage.

\bibliographystyle{IEEEtran}
\bibliography{references}
\end{document}